\documentclass[10pt,number,amsmath,amssymb]{elsarticle}
\biboptions{sort&compress}
\usepackage{graphicx}
\usepackage{fancyhdr}   
\usepackage{amsmath}    
\usepackage{comment}    
\usepackage{caption}
\usepackage{fancybox,shadow}
\usepackage{morefloats}

\newcommand{\be}{\begin{eqnarray}}
\newcommand{\ee}{\end{eqnarray}}
\usepackage{makeidx}  

\usepackage{cite}
\usepackage{url}

\usepackage{amsfonts}
\usepackage{xspace}
\date{}

\title{{\small Reply to comment}\\The Reasonable Effectiveness of Agent-Based Simulations in Evolutionary Game Theory\\Reply to comments on ``Evolutionary Game Theory using Agent-based Methods"}
\author[mmg,pa,bea]{Christoph Adami\corref{cor1}}
\ead{adami@msu.edu}

\author[cse,bea]{Jory Schossau}
\ead{jory@msu.edu}

\author[ib,cse,bea]{Arend Hintze}
\ead{hintze@msu.edu}

\cortext[cor1]{Corresponding author}

\address[mmg]{Department of Microbiology and Molecular Genetics}
\address[pa]{Department of Physics and Astronomy}
\address[cse]{Department of Computer Science and Engineering}
\address[ib]{Department of Integrative Biology}
\address[bea]{BEACON Center for the Study of Evolution in Action,\\ Michigan State University, East Lansing, Michigan, USA}

\begin{document}
\begin{keyword}
Evolutionary game theory \sep Agent-based modeling
\end{keyword}
\maketitle

In 1960, the eminent physicist Eugene Wigner published an essay entitled ``The Unreasonable Effectiveness of Mathematics in the Natural Sciences"~\citep{Wigner1960}, in which he mused about the relationship between physics and mathematics. Wigner expressed his surprise at how effective mathematics is in predicting the future physical world, given certain initial conditions. To him, it seemed almost like a miracle that applying the rules of mathematics over and over would not lead the practitioner into a ``morass of contradictions". He found it striking that the world of mathematics has, at first glance, nothing to do with the physical world. Mathematics permits concepts so abstract that they defy our imagination: they are only limited by the mathematician's ingenuity. To think that concepts such as complex numbers, for example, might be applicable to our physical world in which everything is countable and real might seem ludicrous, until we discover that such concepts are essential in understanding quantum mechanics. With that insight, we now know, comes breathtaking predictability. Whenever we think that the chasm between an ever-expanding body of mathematics and our physical theories is widening, a new development narrows it yet again, for example by realizing that an abstruse mathematical invention like ``Morse theory" turns out to be essential to understand complicated concepts in theoretical physics such as supersymmetry~\citep{Witten2014}. Perhaps mathematics is even more unreasonably effective than even Wigner dared to imagine? 

Mathematics truly is the language of nature as Galileo wrote, and the present authors do not believe that this is an accident that must be taken on faith. While mathematics is a system based on logic and consistency, natural laws are based on {\em causality} and consistency. Now, logical inference and causality are not the same thing, but in a computational view of physical law (nature ``computes" the outcomes of experiments) they become one and the same~\citep{Zuse1969} (see also~\citep{Schmidhuber1997}). Yet it may seem like a surprise that physical theories can describe and predict so much. After all, each and every physical situation is complicated by a myriad of initial conditions, of which very few end up being important for prediction. In other words, a theory that describes so much while assuming so little is certainly something special, in particular because it predicts outcomes that by no means were inherent in the initial conditions. Indeed, therein lies the power of fundamental theories: you get vastly more out of them than what you put in.

This detour about the unreasonable effectiveness of mathematics in describing the physical world brings us to the subject matter at hand, namely the question ``What is the place of agent-based simulations in Evolutionary Game Theory?",   along with a discussion of the effectiveness of such simulations. As the target article~\citep{Adamietal2016} points out,  Evolutionary Game Theory (EGT) historically was formulated in purely mathematical terms, and the theory was successful in the sense that it formalized concepts--such as the evolutionarily stable strategies (ESS)--that before their mathematical formulation appeared paradoxical. Of course, this is the strength of the mathematical approach: by casting our thoughts into algebraic terms, we force our thoughts to be logical, and we are led to a deeper understanding simply because our flawed intuitions are superseded by the implacable logic and consistency of mathematics. But we may go further still: By consistently applying the rules of logic to the edifice we created, we can be led to insights that were nowhere apparent when the groundwork was laid. Thus, we obtain more than we put in.

Yet, mathematics as it is today cannot describe all of nature. Wigner in his essay wondered whether a  ``theory of biology" must necessarily be consistent with theories of physics. In particular, one of the defining features of physics, namely that physical laws depend on only very few variables (a stone falls according to the same law and in the same way--never mind when, never mind where, never mind on Earth or the moon or anywhere in the universe) seems not to hold for biological systems, where everything seems to depend on everything else. This dependence on an enormous number of initial conditions, as opposed to only a handful, makes it difficult to cast (shoehorn is used more often) biological systems into mathematical terms. Or, we should say, more difficult to describe biological systems in such a way that their future can be predicted as accurately as physics can predict physical systems.

When microscopic algebraic descriptions become unsolvable, computational simulations can take over. There is no ``theory of weather", for example, but the computational simulations of a vast number of particles under the influence of many forces and variables allows us to predict future states from past states, albeit with great effort and with limited accuracy. Nobody would imagine that a simple formula will tell us whether it will rain tomorrow at noon. Are computational simulations going beyond mathematics~\citep{Adami2012}? This question has a different answer depending on how narrow once conceives of mathematics. Inside the computer, variables are updated in a logical and consistent manner, and--barring a mistake--the mechanism will never create ``a morass of contradictions". (Of course, the same can be said of mathematics.) 

As opposed to mathematics, a computer {\em can} create predictions that have nothing to do with nature (and thus end up in a morass of contradictions with the world). After all, it is the programmer who implements the rules of progression (the model), and these may be well-chosen, or poorly. To validate the model, it must be checked across a range of parameters, making contact with known results (often known via mathematics) at the extreme values of the ranges. At the same time, the simulation results must be checked for internal consistency, by creating an ensemble of ``trajectories" via hundreds or thousand of replicate simulations.

To some, computational simulations are but an extension of the mathematician's tool set, exerting control over more variables than our consciousness (or our piece of paper) can keep track of. To others, mathematics must display the relationship between variables openly: just ``cranking the machine forward" to obtain a result is insufficient, as it may lack the insight that a closed-form solution provides. We believe there are truths in both points of views, and most of the comments to our article propose that both approaches--the formal algebraic, as well as the computational--are used side-by-side, so as to harness the power of both worlds.

Schuster~\citep{Schuster2016} in his contribution makes the link between evolutionary game theory and population genetics more apparent. Indeed, there can be no doubt that EGT should be seen as a type of ``effective theory" of biological evolution. Mathematical population genetics is often described as ``the theory of evolution", but just as microscopic theories in physics, population genetics can only describe simplified scenarios, such as the evolution of one or several loci (traits) at the time. Schuster casts EGT in terms of the population genetics of $n$ loci, where each locus determines one strategy, and re-derives the differential equations of game dynamics using this approach. He correctly points out that such an approach (which incidentally is isomorphic to the system of equations pioneered by Eigen and Schuster~\citep{Eigen1971,EigenSchuster1979}), cannot describe what is special about evolution, namely the emergence of new traits as well as the extinction of existing ones. Here again we note that in a limited theory, you do not extract more from the theory than what you put in. However, this formulation has several  advantages over standard population genetics. For one, the theory describes evolutionary dynamics at arbitrary mutation rates, as opposed to population genetics that is essentially the zero-mutation-rate limit of Eigen-Schuster theory~\citep{Wilke2005}. Second, frequency-dependent selection is built-in, while in the standard Crow-Kimura formulation of population genetics~\citep{CrowKimura1970} the fitness landscape is constant and independent of genotype frequencies. As a matter of fact, in our review we point out that agent-based simulations of stochastic strategies at finite mutation rate confirm that the strategies that dominate (indeed, are evolutionarily stable) are not monomorphic, but rather are stationary mixtures of strategies that correspond precisely to the quasi-species solutions of Eigen and Schuster. 

In the end, this should not have surprised us. The replicator-mutator equations of EGT are effectively the equations of Eigen and Schuster, and therefore have the same solutions at finite mutation rate. But Schuster points out that analytical solutions to these equations are only possible for very simple fitness landscapes. To go beyond those, he writes that ``agent-based modeling is the method of choice", but that nevertheless important insights can be gained from the simple limiting cases. 

Tarnita in her comment~\citep{Tarnita2016} makes a point worth repeating: just because the mathematical tools available today are unable to describe a good fraction of the biologically relevant scenarios in EGT (most notably, significant rates of mutation, strong selection, and spatial dynamics), this does not mean that those tools will never be available, and one should not stop attempts at developing them just because agent-based simulations are successfully tackling those situations today. Indeed, Tarnita has developed clever methods to deal with the strong mutation regime, albeit in the limit where mutation rates are so large that all strategies are equally likely (which is essentially the mutational meltdown regime~\citep{Antaletal2009,Tarnitaetal2009}). While it is possible to describe intermediate mutation rates via an approximative interpolating scheme, the approach does not capture the intricate interplay between  the stochastic strategies that is the hallmark of quasistrategies described in the target article. But to echo the comment that mathematical approaches should not be disregarded, much more mathematical work is required to understand the nature of the quasistrategy, as it requires solving the Fokker-Planck equation for the strategy distribution function in multiple dimensions~\citep{AdamiHintze2016}.

Hilbe and Traulsen in their comment~\citep{HilbeTraulsen2016} marshal the now well-known backstory of how the ZD strategies of Press and Dyson~\citep{PressDyson2012} were discovered, to illustrate what should perhaps be the most important ``take-home-message" of the target article as well as the comments: Mathematics and agent-based simulations should be used side-by-side by anyone who is interested in making progress in the fascinating field of evolutionary game theory. In the case of the ZD strategies, they were discovered after a fully numerical sweep turned out peculiar regularities that cried out for an algebraic understanding. This is not the first time that results from a computer simulation have hinted at underlying mathematical structures, nor will it be the last. The understanding, of course, was provided by mathematics (in particular, Freeman Dyson's, who is a mathematician turned physicist). But for all the mathematical understanding, the evolutionary fate of ZD strategies is not currently amenable to closed-form solution, even though it is not inconceivable that they may one day be. The agent-based simulations of ZD strategy evolution in the target article revealed that at modest mutation rates, the mathematically favored ``generous" ZD strategies lose out to robust quasistrategies that are {\em not} of the ZD type, because they are groups of strategies that support each other via mutations: the precise analogue of the quasispecies of Eigen and Schuster. 

As Hilbe and Traulsen point out, using agent-based simulations for discovery can be a dangerous undertaking. The end result of a simulation is a set of numbers that were obtained from another set of numbers that came from other numbers, through a sometimes million-fold iteration. How do you protect yourself from random elements affecting your end result? Perhaps surprisingly, you do this as you do in any other scientific endeavor. When deriving mathematical results (sometimes with hundreds if not thousands of transformations), you must constantly check against known results. Often, in a complex mathematical formula a limit can be taken that should recover a known result. Researchers often go through such ``sanity checks" to make sure that no error has crept in while deriving new results that cannot be checked, simply because the results are new. In long series of molecular lab bench experiments, practitioners also have checkpoints built in, as well as controls, that allow them to catch mistakes. Computational science is no different in this respect. Negative controls must be run (as well as sometimes positive controls, if possible), and ``sanity checks" must be provided by examining limits where mathematical results are available. So, for example, when we determined that robust quasistrategies can outperform the generous strategies of Stewart and Plotkin~\citep{StewartPlotkin2013} in the strong mutation regime, naturally we checked that if the mutation rate was brought into the weak mutation regime, the generous strategies became victorious, as mathematics had predicted they would.
So without a doubt, agent-based simulations should never be the only approach taken. But not using this tool because results do not immediately provide intuition, or because one can't be sure that errors have crept in, would be akin to disregarding the results of any experiment conducted anywhere. 

Finally, the comment of Bellomo and Elaiw~\citep{Bellomo2016} fleshes out the discussion of spatial interactions in more detail than we gave it in the target article. When interactions between agents are simple, mathematical descriptions of populations can be used, and in the target article we discuss several such applications. Bellomo and Elaiw highlight in particular applications of agent-based simulations in social economics as well as in crowd behavior, where the interactions between agents can be complex, involve memories of past encounters, have a stochastic element, are non-transitive, or even nonlinear. Such agent-based simulations can lead to surprising insights (see, e.g., the simulation of escape panic behavior~\citep{Helbingetal2000}). After all, this must be measuring stick for using agent-based simulations: if you cannot get more out of them than you put in, then you are not using this tool optimally. 

Because agent-based simulations rely on the applications of defined rules in a systematic and consistent fashion, the success of agent-based simulations in EGT is not at all unreasonable. It is in fact expected as long as such simulations are used with the same kind of care and control that the theoretical and experimental approach is known for. It is true that the ``science of computational simulation" is not always taught, and because it is a relative newcomer in the arsenal of scientific inquiry, it is often neglected. It is perhaps because such caution is not always exercised that computational simulations (in particular of the evolutionary process) are sometimes greeted with skepticism. But giving up on them would turn out to be a missed opportunity rivaling that which Dyson wrote about more than 40 years ago, when he lamented the missed opportunities of mathematicians that could have drawn inspiration from modern developments in theoretical physics~\citep{Dyson1972}. The estrangement between mathematics and physics that Dyson bemoaned appears to be a thing of the past, but we should be ever watchful that disciplines that can inform each other (here, mathematics and computational science) do not go their separate ways. Thus we would like to close this response with the same call that Dyson issued in his essay, who quoted the mathematician Jacques Hadamard who warned (in the gender-specific way of those times):

``It is important for him who wants to discover not to confine himself to one chapter of science, but to keep in touch with various others."
\section*{Acknowledgements}
This work was supported in part by Michigan State University through computational resources provided by the Institute for Cyber-Enabled Research. The material is based in part on work supported by the National Science Foundation under Cooperative Agreement No. DBI-0939454.

\end{document}